\documentclass{article}
\usepackage{spconf,amsmath,amssymb,graphicx, url, multirow}
\usepackage{color,soul}

\def\BLEU{{BLEU}}
\def\BLEUone{{BLEU\textsubscript{1}}}
\def\BLEUtwo{{BLEU\textsubscript{2}}}
\def\BLEUthree{{BLEU\textsubscript{3}}}
\def\BLEUfour{{BLEU\textsubscript{4}}}
\def\ROUGEL{{ROUGE\textsubscript{L}}}
\def\METEOR{{METEOR}}
\def\CIDER{{CIDEr}}

\title{Automated Audio Captioning with Recurrent Neural Networks}
\name{Konstantinos Drossos\textsuperscript{*}, Sharath Adavanne\textsuperscript{*}, and Tuomas Virtanen\thanks{\textsuperscript{*}Equally contributing authors in this paper.}\thanks{The research leading to these results has received funding from the European Research Council under the European Union’s H2020 Framework Programme through ERC Grant Agreement 637422 EVERYSOUND.}\thanks{Part of the computations leading to these results were performed on a TITAN-X GPU donated by NVIDIA. The authors also wish to acknowledge CSC-IT Center for Science, Finland, for computational resources.}\thanks{The authors would like to thank S. Venugopalan (U.T.A., U.S.A.), S. Mimilakis (Fraunhofer IDMT, Germany), and A. Mesaros (TUT, Finland) for the fruitful conversations and valuable feedback.}}
\address{Tampere University of Technology, Tampere, Finland}

\begin{document}
%
\maketitle
\begin{abstract}
We present the first approach to automated audio captioning. We employ an encoder-decoder scheme with an alignment model in between. The input to the encoder is a sequence of log mel-band energies calculated from an audio file, while the output is a sequence of words, i.e. a caption. The encoder is a multi-layered, bi-directional gated recurrent unit (GRU) and the decoder a multi-layered GRU with a classification layer connected to the last GRU of the decoder. The classification layer and the alignment model are fully connected layers with shared weights between timesteps. The proposed method is evaluated using data drawn from a commercial sound effects library, ProSound Effects. The resulting captions were rated through metrics utilized in machine translation and image captioning fields. Results from metrics show that the proposed method can predict words appearing in the original caption, but not always correctly ordered.
\end{abstract}
\section{Introduction}
\label{sec:intro}	
The automated audio captioning problem can be defined as the task of automatically generating a textual description (i.e. caption) for an audio signal, where the caption is as close as possible to a human-assigned one for the same audio signal. This is different from the sound event detection (SED) and audio tagging~\cite{adavanne2017, wichern:2010:icassp}, because the audio captioning method does not predict sound events and their start and end times (as in SED) nor assigns labels to an audio file or parts of it (as in audio tagging). For example, a method for audio captioning must be able to generate descriptions like ``large flock off and flying away'', ``footsteps down on wooden stairs, two men slow approach'', and ``domestic clock striking five''~\footnote{These captions are from PSE Library~\cite{pse:2015:pse}}. 

In other domains, the automated captioning task has previously been explored with different types of multimedia content. In particular, automated image captioning~\cite{vinyals:2015:showandtell, you:2016:cvpr} can be considered as the first attempt to automatically create descriptions for multimedia data, followed by automated video captioning~\cite{subhashini:2015:naacl, shin:2016:icip}. All the above-cited works of automated captioning employ a scheme which is also seen in machine translation field~\cite{cho:2014:emnlp, bahdanau:2015:iclr} and in the zero-shot translation system by Google~\cite{wu:2016:bridging, johnson:2016:zeroshot}. In fact, some image captioning works were inspired by or based on existing works in machine translation, e.g.~\cite{vinyals:2015:showandtell}. In this scheme, the methods can be divided into two parts, the encoder and the decoder. The encoder processes the input data and creates a higher level and/or rich representation of them~\cite{vinyals:2015:showandtell}. The decoder takes as input the output of the encoder and generates the final output sequence, i.e. translated version of the input for the machine translation or caption(s) for the automated captioning task. Some works implement the encoder as a multi-layered convolutional neural network (CNN)~\cite{vinyals:2015:showandtell, subhashini:2015:naacl} and others as multi-layered RNN~\cite{cho:2014:emnlp, bahdanau:2015:iclr}. The decoder is usually implemented as an RNN (multi-layer or single-layer) with a fully connected or maxout~\cite{goodfellow:2013:maxout} layer on top~\cite{cho:2014:emnlp, bahdanau:2015:iclr, vinyals:2015:showandtell}

In this work, we present the first attempt for automated audio captioning. Given an audio file, we seek to generate a textual description that can describe that audio file and is as close as possible to human-assigned captions for the same audio file. To do so, we employ an encoder that processes extracted features from the audio file and feeds it to a decoder, which in turn generates the caption for the audio file word-by-word. The network is trained using audio and captions from a commercial sound effects library, the ProSound effects (PSE) library~\cite{pse:2015:pse}, and we evaluate the produced captions with the metrics employed in Microsoft COCO Caption evaluation~\cite{chen:2015:microsoftcc}, namely \BLEU~\cite{papineni:2002:bleu}, \ROUGEL~\cite{lin:2004:rouge}, \METEOR~\cite{lavie:2007:meteor}, and \CIDER~\cite{vedantam:2015:cider}. The rest of the paper is organized as follows. The proposed method is presented in Section~\ref{sec:method}. The evaluation procedure, obtained results and discussions are presented in Section~\ref{sec:evaluation}. The conclusions and the future expansions for the proposed audio captioning task is discussed in Section~\ref{sec:conclusion}.

\section{Proposed method}\label{sec:method}
Our proposed method takes as an input an audio file of standard CD quality (i.e. 44.1 kHz sampling frequency and 16 bits sample width) and creates a textual description (i.e. caption) for it. At first a matrix of features, $\mathbf{X}\in\mathbb{R}^{T\times N_{\text{feats}}}$, is extracted from the audio file. Each row of the matrix $\mathbf{X}$ contains the extracted features from the input audio file in frame $t$. The matrix $\mathbf{X}$ is used as an input to a neural network, which outputs a matrix $\mathbf{Y}\in\mathbb{R}^{I\times N_{\text{words}}}$. Each each row of the output matrix $\mathbf{Y}$ contains the probability distribution over unique words. Then, we pick the most probable word according to each row of the output matrix, creating a sequence of words. The resulting sequence of words is the caption of the input audio file. 

For feature extraction, the input audio file is segmented using Hamming window of 46 ms (2048 samples) and 50\% overlap. From each resulting frame we extract $N_{\text{feats}}=64$ log mel-band energies. With the extracted audio features, we create the matrix $\mathbf{X}=[\mathbf{x}_1, \ldots, \mathbf{x}_{T}]$, where $\mathbf{x_t}\in\mathbb{R}^{64}$ is a vector containing the log mel-band energies in frame $t$, and $T$ is the number of frames that the input audio data is divided into.

The neural network consists of an encoder, a soft alignment model, and a decoder. The encoder is a three-layered bi-directional GRU with tanh output activations and a residual connection between the second and the third layer. All but the last layer have 64 cells and the last one 128. This results to 128 and 256 cells due to bi-directionality ($2\times64$ and $2\times128$), respectively. The output $\mathbf{h}^{l}_{t}$ of the bi-directional GRU layer $l$ is, according to~\cite{bahdanau:2015:iclr}, 
\begin{equation}\label{eq:bidir_hidden_output}
\mathbf{h}^{l}_{t} = [{\overrightarrow{\mathbf{h}^{l}_t}^T}; {\overleftarrow{\mathbf{h}^{l}_t}^T}]^T\text{,}
\end{equation}
\noindent
where $\overrightarrow{\mathbf{h}^{l}_{t}}$ is the hidden output of the forward GRU layer $l$, $\overleftarrow{\mathbf{h}_{t}}$ of the backward one. The encoder takes as input the matrix $\mathbf{X}$ and produces as an output the matrix $\mathbf{H}^3=[\mathbf{h}^3_1,\ldots,\mathbf{h}^3_T]$, with $\mathbf{h}^3_t\in\mathbb{R}^{256}$.

The soft alignment model is a fully-connected layer with softmax activation. It takes as input the output of the encoder (i.e. $\mathbf{H}^3$) and the hidden recurrent output $\mathbf{h}'_{i-1}$ of the decoder, and produces as an output the $i$-th input to the decoder, $\mathbf{c}_i$. The soft alignment model helps the decoder focus on different parts of the output of the encoder (i.e. weight differently each $\mathbf{h}^3_t$) for his $i$-th prediction, while incorporating information from the previous states and predictions of the decoder (due to $\mathbf{h}'_{i-1}$). For that reason, and similar to~\cite{bahdanau:2015:iclr}, we interpret the alignment model as a soft attention mechanism and the resulting weighted sum, $\mathbf{c}_i$, as a context vector. The context vector $\mathbf{c}_i$ is the weighted sum of the output sequence of the encoder and, according to~\cite{bahdanau:2015:iclr}, calculated as
\begin{equation}
\mathbf{c}_i = \sum\limits^T_{t=1}Fully Connected(\mathbf{h}^3_{t}, \mathbf{h}'_{i-1})\times\mathbf{h}^3_{t}\text{, }
\end{equation}
\noindent
where the $Fully Connected$ is the fully-connected layer of the alignment model with softmax activation which outputs a scalar in the range of (0, 1), $\mathbf{h}'_{i-1}$ is the hidden recurrent output of first layer of the decoder for the $i-1$ prediction, and $\mathbf{h}_{0}$ is a vector of zeros. The soft alignment model has shared weights between calculations of the different $\mathbf{c}_i$.

The decoder is a two-layered GRU with tanh activation functions, followed by a fully-connected layer with softmax activation. The first GRU layer of the decoder has 128 cells and the second 256. It takes as input the outputs of the soft alignment model and produces the matrix $\mathbf{Y}=[\mathbf{y}_1, \ldots, \mathbf{y}_{I}]$, where $\mathbf{y}_i\in\mathbb{R}^{N_{\text{words}}}$ is a vector containing the probability distribution over $N_{\text{words}}$ unique words. From each $\mathbf{y}_i$, we select the word with the highest probability. This results at a sequence of $I$ words. This sequence of $I$ words is the predicted caption for the input audio file. 

\begin{figure}
  \centering
  \includegraphics[width=.83\columnwidth]{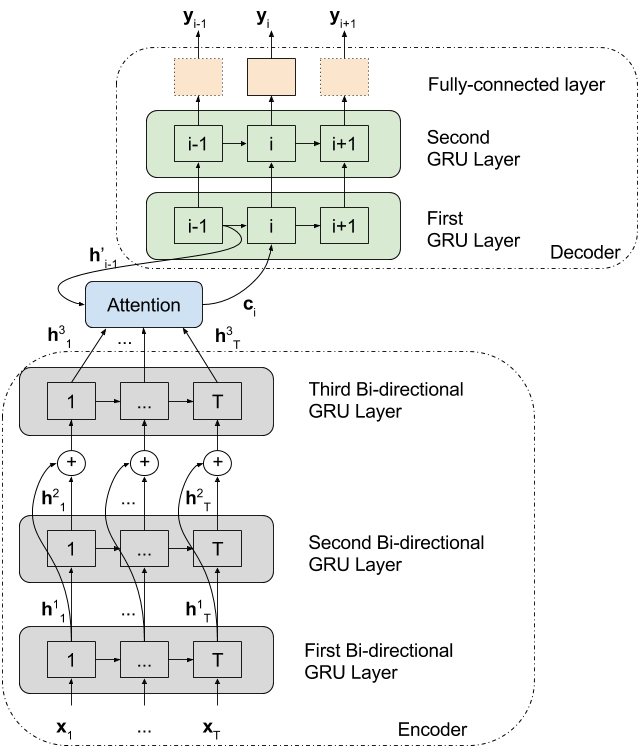}
  \caption{Illustration of the presented neural network.}
  \label{fig:arch}
  \vspace{-5pt}
\end{figure}

An illustration of the neural network is in Figure~\ref{fig:arch}. The amount of the total parameters of the network is 928 638. The encoder, the attention mechanism, and the decoder are jointly trained using Adam optimizer with the default parameters mentioned in the original paper~\cite{kingma:2015:adam}, and categorical-crossentropy loss. Dropout rate is 0.5 and 0.25 for the input and recurrent connections, respectively, in the GRUs of the encoder and the decoder, and the batch size is 64. All the above mentioned hyper-parameters, including the amount of layers for the encoder and decoder, are set after a grid search, using validation data. The initializations of the weights of the GRU inputs and the fully-connected layers are performed using Glorot \& Bengio uniform initialization~\cite{glorot:2010:init}. The weights of the recurrent connections of the GRU layers are initialized using orthogonal initialization. The code is developed using the keras~\cite{chollet:2015:keras} framework with Theano backend~\cite{theano:2016}. The attention mechanism is implemented by the authors according to~\cite{bahdanau:2015:iclr}.

\section{Evaluation}\label{sec:evaluation}
\subsection{Dataset pre-processing}\label{subsec:audioprocessing}
In order to evaluate our method, we employ the audio data and their corresponding captions from PSE library~\cite{pse:2015:pse}. This library consists of 175,000 sound recordings distributed over 501 categories in total, e.g. ambience, audience, foley, nature, wildlife, sports, transportation and weather. The audio recordings in the PSE library are inconsistent in terms of sampling rate, number of channels, and sample width. In order to have consistency in the data we used recordings with sampling rate of at least 44.1 kHz, maximum two channels, and sample width of at least 16 bits. This resulted in 153 220 recordings. 

All 153 220 recordings were processed to have a sampling rate of 44.1 kHz, sample width of 16 bits, and single channel. Each of the recordings is associated with one textual description (i.e. caption). The captions in most cases are not proper sentences but rather a set of keywords. Within a caption some keywords repeat more than once and in different forms (e.g. plural and singular). The repeating words in the same form were removed (i.e. from ``bird bird'' we removed only one word resulting in ``bird''). Removal of punctuations was performed in order to reduced the final amount of unique tokens/words that have to be predicted. The words in captions have also a plenty of typographic/orthographic errors. This results in a very high number of unique words. In order to overcome this, we removed all words which are not in the US or UK English dictionary using Enchant python package~\cite{enchant:python} and GNU-Aspell dictionaries~\cite{gnu:aspell}. This reduced the unique words from 19,638 to 11,478 and the total number of words from 1.3 M to 1.16 M. This also resulted in 601 captions with zero words that were excluded from our experiments. This resulted in 152 619 recordings, each having caption length from one to 60 words. 

We use a randomly chosen subset of approximately 10\% of the 152 619 cleaned library as our data set. This subset is further randomly split to approximately 60\% training, 20\% validation, and 20\% testing data. This corresponds to 8757, 2975 and 2960 recordings, respectively, and amounts to 14,692 recordings. Some captions of this 10\% subset are identical between different recordings. In order to ensure that our proposed method does not memorize captions or similar acoustic conditions (implied by the similarity of the captions), we create the data test split with recordings that have captions which do not occur in either training or validation split. In the data splits there might be cases where the captions in the test split contain words not appearing in the training and validation splits. This would result in trying to predict unknown words (i.e. words not appearing in the training split). Given also that in the set of 14,692 recordings $75.36\%$ of words appeared less than five times, the probability of having words in the test split that are not appearing in the training one, increased. We tackle this by creating 1000 random sets of splits (i.e. training, validation, and testing splits is one set) fulfilling the restriction of not having identical captions in the three splits. From these 1000 sets, we choose the one that had the maximum amount of common words, with the restriction that these words also occurred at least five times in each of the splits. The chosen set of splits contained a total of 71000 words with $N_{\text{words}}=633$ unique words. Finally, we chose a caption length of maximum 10 words followed by an end of sentence (\textless EOS\textgreater) token as the caption for each of the recordings. This results to $I=11$. Captions longer than 10 words were trimmed, and shorter were padded with \textless EOS\textgreater{} token. The length was motivated from the dataset, where approximately $90\%$ of the recordings (13220 out of 14692) had captions equal to or less than 10 words. We utilize only the first thirty seconds of the audio of each recording. Recordings shorter than this were padded with zeros. This results to $T=1289$ frames of audio.

\subsection{Evaluation metrics and procedure}
The proposed method was evaluated according to the following metrics used in the machine translation and image captioning fields. These metrics were calculated for 10 different training and testing runs, i.e. for each run the parameters of the neural network were re-initialized according to the initialization functions and the neural network was re-trained on the training split.

\BLEU~\cite{papineni:2002:bleu} is a precision-based metric. It calculates a weighted geometric mean of a modified precision of $n$-grams between predicted and ground truth captions. Due to the calculation of the modified precision that favors short predicted captions, \BLEU~uses a brevity penalty in the calculation of the geometric mean. This penalty penalizes predicted captions that are shorter than the ground truth ones. Typical lengths for $n$-grams are one to four, resulting in $\text{BLEU\textsubscript{1/2/3/4}}$, respectively~\cite{papineni:2002:bleu, chen:2015:microsoftcc}. \METEOR~\cite{lavie:2007:meteor} calculates a harmonic mean of precision and recall of segments of the captions between the predicted and ground truth captions. The recall is weighted significantly more than precision and thus \METEOR~is considered a recall-based metric~\cite{chen:2015:microsoftcc}. It employs alignment between the words of the predicted and ground truth captions and matches exact words, stems of words, synonyms, and paraphrases. The alignment is computed over segments of the captions (chunks) between the ground truth and predicted captions while also minimizing a number of chunks needed~\cite{lavie:2007:meteor, chen:2015:microsoftcc}. \CIDER~\cite{vedantam:2015:cider} calculates a weighted sum of the cosine similarity between the predicted and ground truth captions for $n$-grams of length $n$ with $n\in [1, 4]$. The cosine similarity is calculated using Term Frequency Inverse Document Frequency (TF-IDF) weighting for each $n$-gram~\cite{vedantam:2015:cider, chen:2015:microsoftcc}. The \CIDER~that is employed to MS COCO Caption evaluation (and is used in this work) is a modified version of \CIDER~that is more robust to gaming and is called CIDEr-D. Gaming refers to the fact that sentences which scored high with an automated metric, tend to be evaluated poorly by humans~\cite{chen:2015:microsoftcc}. \ROUGEL~\cite{lin:2004:rouge} is a Longest Common Subsequence (LCS) based metric. It calculates an F-measure using LCS between the predicted and ground truth caption. The F-measure is oriented towards recall using a value for the $\beta=1.2$ in the F-measure calculation~\cite{lin:2004:rouge, chen:2015:microsoftcc}.

Since the current paper presents the very first method for audio captioning, there are no previous results to compare the presented ones. For that reason, we employed two additional evaluation cases. The predicted captions for the first case are generated by randomly picking one to ten words from the set of unique words in the dataset. This case will show the values for the metrics obtained by random predictions of words. Given the lack of previous studies on automated audio captioning, we use these values of the metrics as indicative lower threshold for the performance of our proposed method. The captions of the second case are generated from testing the network with random input data instead of extracted audio features, along with the ground truth captions. The random input data is a matrix with same dimensions as $\mathbf{X}$ and with values randomly drawn from the distribution of $\mathbf{X}$ for the data in the training split. We use random input data case in order to investigate if the neural network has learned to process the input data in a useful manner and discover relationships between patters of the input data and the patterns of the desired output. 
 
\begin{table}
\centering
\caption{Obtained results for the evaluation metrics}
\label{tab:results}
\scalebox{.95}{
\begin{tabular}{l|llc}
\multirow{2}{*}{\textbf{Metric}} & \multicolumn{3}{c}{\textbf{Case}} \\ \cline{2-4} 
 & \multicolumn{1}{c|}{\textbf{\begin{tabular}[c]{@{}c@{}}Random\\ words\end{tabular}}} & \multicolumn{1}{c|}{\textbf{\begin{tabular}[c]{@{}c@{}}Random\\ input data\end{tabular}}} & \textbf{\begin{tabular}[c]{@{}c@{}}Proposed\\ method\end{tabular}} \\ \hline
\BLEUone & \multicolumn{1}{l|}{0.003 $\pm$0.000} & \multicolumn{1}{l|}{0.006 $\pm$0.001} & 0.191 $\pm$0.004 \\
\BLEUtwo & \multicolumn{1}{l|}{0.000 $\pm$0.000} & \multicolumn{1}{l|}{0.002 $\pm$0.000} & 0.129 $\pm$0.003\\
\BLEUthree & \multicolumn{1}{l|}{0.000 $\pm$0.000} & \multicolumn{1}{l|}{0.000 $\pm$0.000} & 0.106 $\pm$0.003 \\
\BLEUfour & \multicolumn{1}{l|}{0.000 $\pm$0.000} & \multicolumn{1}{l|}{0.000 $\pm$0.000} & 0.094 $\pm$0.003 \\
\ROUGEL & \multicolumn{1}{l|}{0.004 $\pm$0.000} & \multicolumn{1}{l|}{0.008 $\pm$0.002} &  0.149 $\pm$0.002\\
\METEOR & \multicolumn{1}{l|}{0.004 $\pm$0.000} & \multicolumn{1}{l|}{0.004 $\pm$0.001} & 0.092 $\pm$0.002 \\
\CIDER & \multicolumn{1}{l|}{0.005 $\pm$0.001} & \multicolumn{1}{l|}{0.026 $\pm$0.012} & 0.526 $\pm$0.012
\end{tabular}
}
\end{table}

\subsection{Results and discussion}\label{sec:results}
Table~\ref{tab:results} presents the obtained results for the three evaluation cases. The results reported are the mean and standard deviation of the 10 runs. As can be seen, for the proposed method the precision-based metric \BLEU~drops with the increase of the $n$-gram length. Comparing it to the recall-based metric \METEOR, it can be seen that the proposed method has better precision than recall. Additionally, according to the fragment penalization that is performed in \METEOR~and the trend of values of \BLEU~scores, it can be inferred that the proposed method tends to correctly produce words that appear in the original caption but not in the right order. This is also supported by the high value of \CIDER~(namely 0.526), the value of \ROUGEL~(namely 0.149), and the fact that most of the produced captions ($56.59\%$) are shorter than the original ones. The case with the random words indicate that the results obtained with our method are well above of a random guess of a sequence of words. The values of the metrics for the random input data case indicate that the proposed neural network has learned to process the input data in a useful manner and discover the relationships between the input audio features and output caption. 

The captions of the employed data set (PSE library) are not syntactically structured. By checking the predicted captions, we also observed that our method tends to generate captions that can describe the sound but in different words. For example, our method predicted ``field dark'' for a recording with an original caption of ``electric hum deep''. This fact brings forward the well-known ambiguity for sound recognition when no other context is been given. This ambiguity was greatly employed in the initial days of audio effects where sound events were represented with similar sounds, e.g. the sound of crinkled thin plastic used to represent the sound of fire. Below are three example predictions of our method that try to demonstrate the above facts and not the performance of the method itself. \textbf{GX} is the ground truth and \textbf{PX} is the corresponding predicted caption. We present more exhaustive examples in the corresponding on-line demo\footnote{\url{http://arg.cs.tut.fi/demo/captioning/}}.

\textbf{G1:}~goose cu flock calls with distant and near end

\textbf{P1:}~cu gull calls medium other birds md birds in

\textbf{G2:}~footsteps wood down interior

\textbf{P2:}~footsteps belt

\textbf{G3:}~door close slide

\textbf{P3:}~squeak squeak

\vspace{-5pt}
\section{Conclusion and future work}
\label{sec:conclusion}
In the present work, we proposed the first method for automated audio captioning. The method is based on an encoder-decoder scheme with an attentional layer in between them. The method was evaluated using a commercial dataset of recordings, each of which is associated with a textual description (caption) within the dataset. The evaluation of the proposed method was performed with machine translation metrics commonly used in image and video captioning tasks. The results showed that our method can identify the information in an audio recording and produce a set of words that can describe the recording to a certain extent, but still far from intelligible caption. 

Recommendations for future work on the problem of audio captioning include the usage of methods that compensate for audio and captions with different lengths, utilization of language model, and the employment of a different data set with more than one caption per recording and captions that exhibit increased syntactic structure. 


\bibliographystyle{IEEEbib}
\bibliography{refs}

\end{document}